\documentclass[twocolumn,superscriptaddress]{revtex4-1}
\usepackage[final]{graphicx}
\usepackage{amsmath}
\usepackage{hyperref}
\usepackage{soul}

\hypersetup{colorlinks=true, citecolor=blue, urlcolor=blue, linkcolor=blue}

\usepackage{color}	
\newcommand{\tred}[1]{\textcolor{black}{#1}}

\begin{document}

\title{Tuning time and energy resolution in time-resolved photoemission spectroscopy \tred{with nonlinear crystals}}

\author{Alexandre Gauthier}
\affiliation{Stanford Institute for Materials and Energy Sciences, SLAC National Accelerator Laboratory, Menlo Park, California 94025, USA}
\affiliation{Geballe Laboratory for Advanced Materials, Departments of Applied Physics and Physics, Stanford University, Stanford, California 94305, USA}

\author{Jonathan A. Sobota}
\affiliation{Stanford Institute for Materials and Energy Sciences, SLAC National Accelerator Laboratory, Menlo Park, California 94025, USA}

\author{Nicolas Gauthier}
\affiliation{Stanford Institute for Materials and Energy Sciences, SLAC National Accelerator Laboratory, Menlo Park, California 94025, USA}
\affiliation{Geballe Laboratory for Advanced Materials, Departments of Applied Physics and Physics, Stanford University, Stanford, California 94305, USA}

\author{Ke-Jun Xu}
\affiliation{Stanford Institute for Materials and Energy Sciences, SLAC National Accelerator Laboratory, Menlo Park, California 94025, USA}
\affiliation{Geballe Laboratory for Advanced Materials, Departments of Applied Physics and Physics, Stanford University, Stanford, California 94305, USA}

\author{Heike Pfau}
\affiliation{Stanford Institute for Materials and Energy Sciences, SLAC National Accelerator Laboratory, Menlo Park, California 94025, USA}
\affiliation{Advanced Light Source, Lawrence Berkeley National Laboratory, Berkeley, California 94720, USA}

\author{Costel R. Rotundu}
\affiliation{Stanford Institute for Materials and Energy Sciences, SLAC National Accelerator Laboratory, Menlo Park, California 94025, USA}

\author{Zhi-Xun Shen}
\affiliation{Stanford Institute for Materials and Energy Sciences, SLAC National Accelerator Laboratory, Menlo Park, California 94025, USA}
\affiliation{Geballe Laboratory for Advanced Materials, Departments of Applied Physics and Physics, Stanford University, Stanford, California 94305, USA}

\author{Patrick S. Kirchmann}
\email{kirchman@stanford.edu}
\affiliation{Stanford Institute for Materials and Energy Sciences, SLAC National Accelerator Laboratory, Menlo Park, California 94025, USA}

\begin{abstract}
    Time- and angle-resolved photoemission spectroscopy is a powerful probe of electronic band structures out of equilibrium. Tuning time and energy resolution to suit a particular scientific question has become an increasingly important experimental  consideration. Many instruments use cascaded frequency doubling in nonlinear crystals to generate the required ultraviolet probe pulses. We demonstrate how calculations clarify the relationship between laser bandwidth and nonlinear crystal thickness contributing to experimental resolutions and place intrinsic limits on the achievable time-bandwidth product. Experimentally, we tune time and energy resolution by varying the thickness of nonlinear $\beta$-BaB$_2$O$_4$ crystals for frequency up-conversion, providing for a flexible experiment design. We achieve time resolutions of 58 to 103~fs and corresponding energy resolutions of 55 to 27~meV. We propose a method to select crystal thickness based on desired experimental resolutions.
\end{abstract}

\date{\today}
\maketitle

\section{Introduction}

Since its inception in the late 1980s by Haight et al. \cite{Haight1988} time- and angle-resolved photoemission spectroscopy (trARPES) with sub-picosecond time resolution has matured into a powerful tool for the study of electronic band structures in nonequilibrium. The principle of this type of pump-probe spectroscopy is well-established \cite{Bovensiepen2012,Smallwood2016,Zhou2018}: an ultrafast light source generates pump and probe pulses at photon energies below and above the sample work function of a few eV, respectively. The pump pulse excites the electrons in the sample, and the probe pulse photoemits from the excited system. By controlling the delay between the two pulses, the transient photoemission intensity is captured. 

In this way, trARPES has been used to measure various physical phenomena of excited solid state systems. This includes studies of unoccupied electronic states above the Fermi level ($E_\textrm{F}$) and the dynamics of photoexcited electron populations therein \cite{Weinelt2004,Perfetti2007,Sobota2012,Wallauer2016,Kuroda2017,Zhang2017,Gierz2017,Na2019,Soifer2019}, gap dynamics of ordered states such as superconductivity \cite{Smallwood2012,Rameau2016,Ishida2016,Parham2017,Boschini2018,Freutel2019} and charge density waves \cite{Schmitt2008,Peterson2011,Mor2017,Monney2018}, binding energy oscillations due to coherent phonons that couple to electronic states \cite{Papalazarou2012,Avigo2013,Yang2019}, and the creation of novel photoinduced phases of matter such as Floquet-Bloch states \cite{Wang2013}.

The wide range of questions naturally implies that requirements for time and energy resolution vary. However, fundamental quantum mechanical uncertainty prevents simultaneous optimization of time and energy resolutions. In most trARPES experiments, energy and time resolution are dominated by the bandwidth and duration of the probe pulse, respectively. All light pulses are subject to an uncertainty relationship between the spectral and temporal intensity of the pulse, known as the Fourier limit. The product of the \tred{standard deviations of the spectral and temporal intensity of a pulse is restricted by the inequality \cite{Rulliere2005}
\begin{equation}
    \label{eq:uncsigma}
    \sigma_E\sigma_t\geq\hbar/2\approx 329 \textrm{~meV~fs}
\end{equation}
with equality only achievable for Gaussian pulses. In practice it is more convenient to characterize pulses by their full widths at half maximum (FWHM). A similar uncertainty relationship can be written for spectral and temporal FWHM, with a lower bound that depends on the specific pulse shape. Most relevant to optics is the inequality for Gaussian pulses \cite{Rulliere2005}}

%full widths at half maximum (FWHM) of the spectral and temporal intensity of a Gaussian pulse is restricted by the inequality \cite{Rulliere2005} 

\begin{equation}
\label{eq:unc} 
\Delta E \Delta t \geq 4\hbar\ln2 \approx 1825 \textrm{~meV~fs}. 
\end{equation}
\noindent  \tred{For a Gaussian pulse to achieve } this lower bound, \tred{it must have} constant phase across all spectral components, i.e. an unchirped pulse with zero group velocity dispersion (GVD) \cite{Trebino2000}. \tred{We note that certain other pulse shapes, such as Lorentzian and hyperbolic secant, exhibit lower limits than that imposed by Eq.~\ref{eq:unc} \cite{Rulliere2005}.} 

\tred{The bound on} time-bandwidth product implies a trade-off between the best achievable time and energy resolution, and sets rigorous bounds on trARPES experiments. For instance, resolving gap dynamics emphasises energy resolution, which needs to be on the order of the gap \cite{Parham2017}. Conversely, capturing electronic scattering processes generally demands superior time resolution \cite{Gierz2017}. A coherent response can only be observed when  energy and time resolution are balanced to permit obtaining commensurate information in both energy and time domain \cite{Yang2019}. The increasing diversity of questions in the field requires optimization of time and energy resolution for a specific problem, and the ability to change the partitioning of the time-bandwidth product efficiently. 

\begin{figure*}[ht!]
\includegraphics[width=\textwidth]{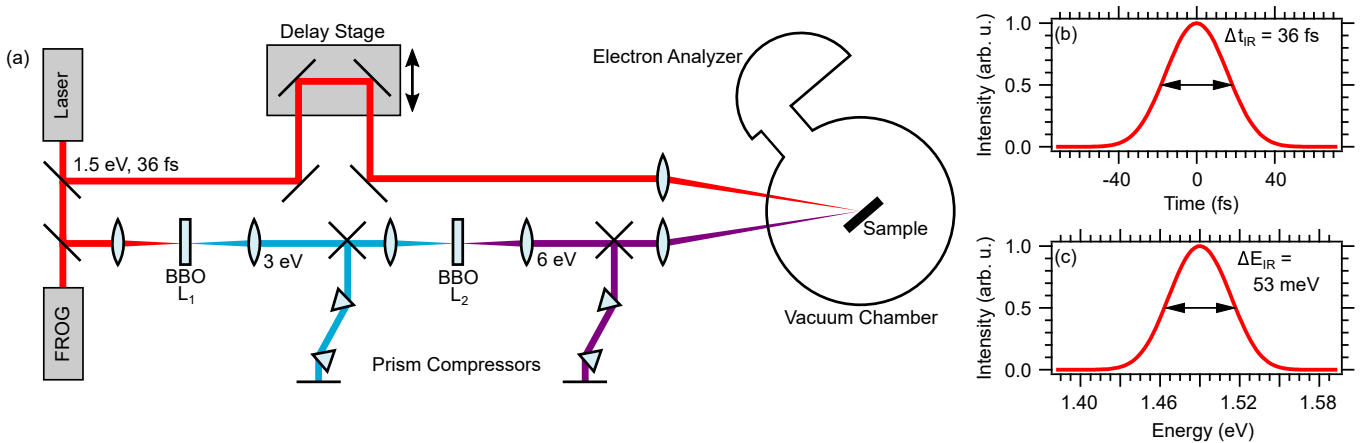}
\caption{(a). Schematic of our time-resolved ARPES laser setup. A 1.5~eV ultrafast laser is split into pump and probe paths. The probe beam is frequency doubled in two nonlinear BBO crystals to reach 6~eV. Prism compressors compensate for propagation through dispersive media \cite{Fork1984}. (b) and (c) show the temporal and spectral intensity respectively of the 1.5~eV pulses at the laser output as measured by a frequency resolved optical grating (FROG) \cite{OShea2001}.}
\label{fig:setup}
\end{figure*}

In order to understand how time and energy resolutions are determined we briefly review how ultrafast probe light pulses for trARPES are generated. The photon energy of the probe pulse must exceed the sample work function for one-photon-photoemission to occur. Typical work function values range from $3-5$~eV but can be as low as 1~eV \cite{Pfau2020} and as high as 6~eV \cite{Sobota2012}. The field has pursued two main avenues to generate ultrafast probe light pulses with $\geq 6$~eV photon energy: high-order harmonic generation (HHG) in gases \cite{Haight1988,Peterson2011,Frietsch2013,Eich2014,Cilento2016,Rohde2016,Wallauer2016,Gierz2015,Gierz2017,Corder2018,Lee2019,Cucini2019,Puppin2019,Buss2019,Mills2019,Sie2019,Roth2019}, and cascaded frequency upconversion in non-linear crystals \cite{Perfetti2007,Schmitt2008,Sobota2012,Smallwood2012-rsi,Faure2012,Boschini2014,Wegkamp2014,Andres2015,Kuroda2017,Ishida2016,Parham2017,Monney2018,Boschini2018,Zhou2018,Freutel2019,Yang2019-rsi}. In addition, the recent development of free electron lasers enables trARPES studies at photon energies in the keV range \cite{Oloff2016,Kutnyakhov2019}.

HHG-based trARPES has the key benefits of working with photon energies $>10$~eV and the ability to select different harmonics to tune the probe photon energy. This permits access to the whole Brillouin zone \cite{He2016} but requires amplified laser systems with pulse energies of $>10$~$\mu$J \cite{Puppin2019}, often considerably more \cite{Frietsch2013}. Such amplifiers tend to work at laser repetition rates below $100$~kHz which can broaden and shift photoemission spectra due to space charge when a large number of electrons are emitted per probe pulse \cite{Hellmann2009}. Notably, recent developments enable considerably higher repetition rates and hence better statistics with less space charge \cite{Hellmann2009,Corder2018,Puppin2019,Mills2019,Nicholson2018}. Borrowing concepts from synchrotron light sources operating at similarly high photon energies \cite{Poletto2007}, the bandwidth of the probe pulse can be adjusted by changing gratings and varying slits \cite{Frietsch2013,Gierz2015,Corder2018,Sie2019,Cucini2019}.

Frequency upconversion in non-linear crystals is limited to photon energies $<7$~eV \cite{Zhou2018} and probe photon energy tuning is limited to a few $100$~meV \cite{Xiong2017}, only allowing studies of the electron dynamics near the Brillouin zone center of most materials. On the upside, generating probe pulses near 6~eV photon energy by quadrupling the fundamental of a Ti:Sa laser operating at 1.5~eV photon energy is considerably easier to implement and more efficient than HHG. Historically, this has enabled higher laser repetition rates with higher statistics trARPES data \cite{Smallwood2016}. It has long been understood that thinner nonlinear crystals for upconversion lead to better time resolution and worse energy resolution \cite{Faure2012}. But a systematic study of rational strategies to achieve a different partitioning of the time-bandwidth product and its implication for trARPES has so far been lacking.

In this paper, we discuss strategies for tuning time and energy resolutions by varying the thickness of nonlinear crystals used to generate probe pulses. We begin in Section \ref{s:Modeling} by surveying the basic considerations for second harmonic generation (SHG) before presenting model calculations. We focus on the most common scheme of two stages of SHG to generate 6~eV probe pulses, and find a delicate interplay between the bandwidth of the laser source and crystal thicknesses contributes to determining time and energy resolutions. Section \ref{s:Experiment} describes our experiments to test different partitions of the time-bandwidth product. With our setup, we demonstrate experimental time resolutions ranging from 58 to 103~fs, with corresponding energy resolutions ranging from 55 to 27~meV, and reach 127\% of the Gaussian Fourier limit. We conclude in Section \ref{s:Discussion} with a comparison of  time-bandwidth parameters together with probe photon energies and laser repetition rates from various groups across the globe, and provide a comprehensive strategy to approach the Fourier limit for a fixed bandwidth of the ultrafast drive laser while maximizing 6~eV photon flux. 

\section{Modeling}\label{s:Modeling}

Here, we discuss the common situation in which a 6~eV probe pulse is generated from a portion of the 1.5~eV fundamental output of a laser system. Taking our Ti:Sa-based amplified laser system as an example, we depict this experimental setup in Fig.~\ref{fig:setup}(a). Two cascaded stages of type I SHG in $\beta$-BaB$_2$O$_4$ (BBO) nonlinear crystals \cite{Kato1986} generate 6~eV probe pulses. The first crystal, of thickness $L_1$, generates 3~eV pulses, and the second crystal, of thickness $L_2$, generates 6~eV pulses. 

\subsection{Basics of Second Harmonic Generation}\label{ss:SHG}

In this section we present the relevant background information on SHG following the formalism of Weiner \cite{Weiner2009}. For simplicity we neglect depletion of the input pulse as it propagates through the crystal \cite{Sidick_1995_ultrashort}, and geometric effects associated with focusing and spatial walk-off \cite{Guha_1980_the,Weiner2009}. Our work suggests that these effects are small when working in common experimental configurations. In Section~\ref{s:Discussion} we will briefly discuss the limitations of our assumptions. Excluding these factors, the input and second harmonic electric fields in the time domain $\mathcal{E}_1(t)$ and $\mathcal{E}_2(t)$ are described as plane waves propagating along the same axis through the crystal. The frequency domain spectra $\tilde{\mathcal{E}}_1(\omega)$ and $\tilde{\mathcal{E}}_2(\omega)$ can be expressed as:
\begin{equation} 
\mathcal{E}_\alpha(t) = \frac{1}{4\pi} \left[ \int_0^\infty \tilde{\mathcal{E}}_\alpha(\omega) e^{i(\omega t - k_\alpha(\omega)z)}d\omega  + \textrm{c.c.}\right] 
\label{eq:field_definition} 
\end{equation}
\noindent where $\alpha\in\{1,2\}$ denotes the first or second harmonic, $k_\alpha=\omega n_\alpha(\omega)/c$ is the wave vector of the respective frequency components, and $n_\alpha$ is the index of refraction of the crystal along the field's polarization vector. After propagating through a nonlinear crystal of length $L$, the second harmonic field is  determined by \cite{Weiner2009}:

\begin{widetext}
\begin{equation}
\tilde{\mathcal{E}}_2(\omega) \propto L\omega\int_0^\infty \tilde{\mathcal{E}}_1(\omega')\tilde{\mathcal{E}}_1(\omega-\omega')e^{i\Delta k(\omega,\omega')L/2}\textrm{sinc}(\Delta k(\omega,\omega')L/2)d\omega'
\label{eq:shg}
\end{equation}
\end{widetext}

\noindent where $\Delta k(\omega,\omega') = k_2(\omega)-k_1(\omega') - k_1(\omega-\omega')$. The integrand is weighted by a $\textrm{sinc}$ function which is peaked at $\Delta k L = 0$. This reflects the phase matching condition in which the frequency components of the second harmonic interfere constructively to maximize the conversion efficiency. In practice, the nonlinear crystal is oriented such that phase matching is achieved ($\Delta k \approx 0$) for the central wavelength of the input pulse, while nearby wavelengths are converted with decreased efficiency. This defines a \textit{phase matching bandwidth} which is the maximum spectral bandwidth of the second harmonic pulses as a function of fundamental wavelength and crystal thickness.

\begin{figure}[ht]
\includegraphics[width=0.8\columnwidth]{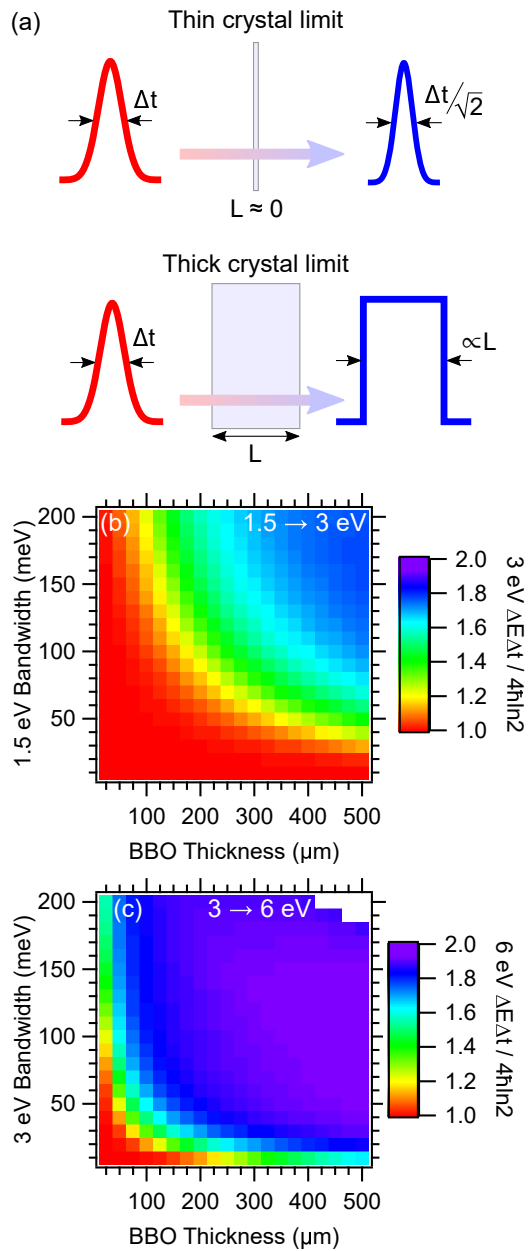}
\caption{(a) Sketch of the limiting cases of nonlinear crystal behavior. For sufficiently thin crystals the pulse duration is reduced by a factor of $\sqrt{2}$, while for thick crystals the pulse becomes rectangular with a duration proportional to $L$ (see Eq.~\ref{eq:shgbw}). (b) and (c) show the time-bandwidth product as a function of BBO thickness and input bandwidth for generation of 3~eV and 6~eV pulses, respectively. Values are scaled such that 1 represents Gaussian pulses.}
\label{fig:shg}
\end{figure}

Some physical intuition can be developed by considering two extreme limits of Eq.~\ref{eq:shg}. In the \textit{thin crystal} limit, the phase matching bandwidth greatly exceeds the bandwidth of the input pulse, giving  $\textrm{sinc}(\Delta k L )\approx 1$ in the integrand. Thus $\tilde{\mathcal{E}}_2(\omega)\propto \int_0^\infty \tilde{\mathcal{E}}_1(\omega')\tilde{\mathcal{E}}_1(\omega-\omega')d\omega'$, with the simple result that $\mathcal{E}_2(t) = \mathcal{E}_1(t)^2$. For a Gaussian input pulse of length $\Delta t$, the SHG pulse will have a length $\Delta t/\sqrt{2}$. This limit explicitly illustrates the second order nature of the process by which components at $\omega'$ and $(\omega-\omega')$ are combined to give a new component at $\omega$. In the \textit{thin crystal} limit, the SHG pulse shape is independent of $L$, and is determined entirely by the input pulse.

In the opposing \textit{thick crystal} limit, the bandwidth of the input pulse greatly exceeds the phase matching bandwidth. It can be shown that $\mathcal{E}_2(t)$ is then a rectangular pulse of duration 
\begin{equation}
\Delta t_\textrm{Rect.}=\Big\lvert \left( \frac{\partial k_2}{\partial \omega} \Big\vert_{2 \omega_0} - \frac{\partial k_1}{\partial \omega} \Big\vert_{\omega_0}  \right) \Big\rvert L,
\label{eq:shgbw}
\end{equation}
\noindent where $\omega_0$ is the phase-matched component of the input pulse \cite{Weiner2009}. In the \textit{thick crystal} limit, the pulse shape of the input pulse is irrelevant, and the SHG pulse shape is entirely determined by the dispersion in and the thickness of the nonlinear crystal. The reciprocal value of Eq.~\ref{eq:shgbw} provides an order-of-magnitude estimate of the phase matching bandwidth of a given crystal. The thin and thick crystal limits are sketched in Fig.~\ref{fig:shg}(a). 

At this point, we can make a general observation about the behavior in these two limiting cases. In the thin crystal limit, Gaussian pulses are converted into Gaussian pulses, making it possible to operate at the Gaussian Fourier limit. In the thick crystal limit, the generation of rectangular pulse shapes necessarily implies a deviation from the Gaussian Fourier limit; rectangular pulses are subject to a different limit $\Delta E\Delta t\geq 3664 \textrm{~meV~fs}$. In intermediate cases, increasing crystal thickness or input bandwidth leads to a deviation from the Gaussian Fourier limit. We shall explore this behavior numerically in the next section, and elaborate  on its experimental significance in the Discussion section.

\subsection{Numerical Calculations of Two-Stage SHG}\label{ss:Calculations}

Now we numerically integrate Eq.~\ref{eq:shg} to understand how the SHG pulse shape evolves between the two limiting cases.  We use values for the refractive index $n(\lambda)$ for BBO published by Tamosauskas \textit{et al.}, \cite{Tamosauskas2018} and compare with coefficients published by Castech \cite{Castech} and Eksma Optics \cite{Eksma} to estimate errors, see Table~\ref{t:compareior} for details. Besides $n(\lambda)$, the only input parameters for the calculation are the BBO thicknesses for second and fourth harmonic generation $L_1$ and $L_2$, respectively, and the measured spectral intensity $\tilde I(\lambda)$ of the fundamental pulse. We restrict the calculations to Gaussian fundamental pulses of varying bandwidth.

First we consider both stages of SHG separately. We evaluate Eq.~\ref{eq:shg} for input pulses at central energies 1.5 and 3~eV, and independently tune the input bandwidths (10 to 200~meV) and BBO thicknesses (25 to 500~$\mu$m). We summarize the results by plotting the time-bandwidth product of the resultant SHG pulses in  Fig.~\ref{fig:shg}(b-c) as a function of both independent variables. The colormap is scaled such that 1 represents the time-bandwidth product of Gaussian pulses. On this scale, the Fourier limit for rectangular pulses is 2.008. Thus, red shaded regions indicate Gaussian pulses, while blue-purple regions indicate that more rectangular pulses are being generated.

Gaussian pulses are only generated for a sufficiently thin BBO or small enough input bandwidth. This corresponds to the \textit{thin crystal} limit where the SHG phase matching bandwidth significantly exceeds input bandwidth. Outside this regime, the input bandwidth exceeds the phase matching bandwidth, and the generated pulses become increasingly rectangular as the \textit{thick crystal} limit is approached. Comparing Fig.~\ref{fig:shg}(b) and (c) reveals that generated 6~eV pulses are Gaussian over a much smaller parameter space than 3~eV pulses. This reflects the fact that the phase matching bandwidth for 6~eV generation for a given crystal thickness is smaller than that for 3~eV generation due to steep material dispersion $\partial n / \partial \omega$ in the ultraviolet (see Eq.~\ref{eq:shgbw}). In practice, this means that in two stage generation of 6~eV pulses from 1.5~eV, the second crystal should be chosen thinner than the first.

Now we consider the conversion process from 1.5~eV to 6~eV pulses as a whole. We vary $L_1$, $L_2$, and $\Delta E_\textrm{IR}$, the bandwidth of the fundamental. The fundamental is centered around 1.5~eV photon energy. The outputs of the calculation are the bandwidth $\Delta E_\textrm{UV}$ and duration $\Delta t_\textrm{UV}$ of the 6~eV pulses, which closely correspond to experimental energy and time resolutions. This discussion highlights general trends without claiming predictive quantitative accuracy for final experimental resolutions, which are also influenced by other factors such as the pump pulse duration and electron spectrometer resolution.

\begin{figure*}
\includegraphics[width=\textwidth]{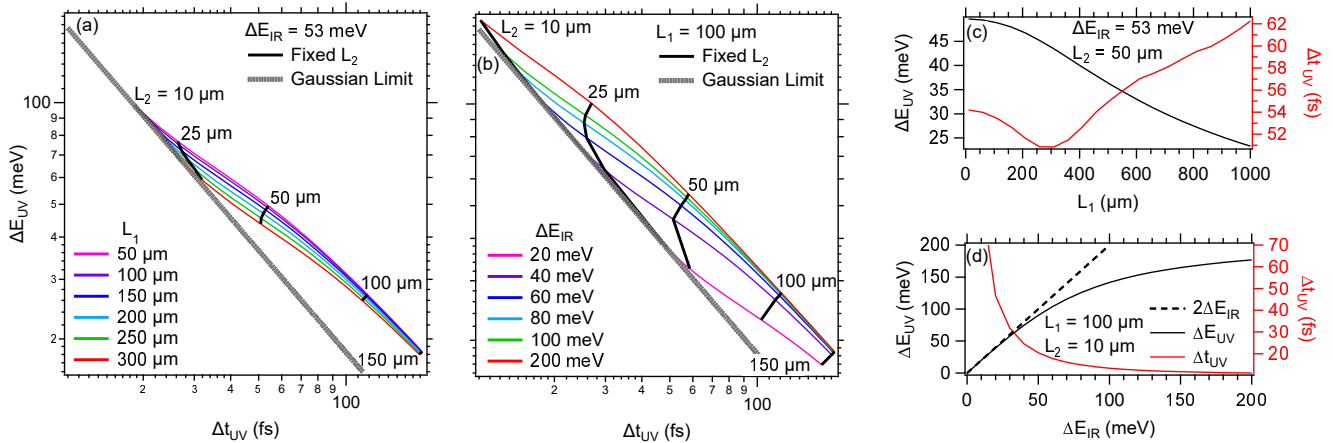}
\caption{(a) Calculated $\Delta t_\textrm{UV}$ and $\Delta E_\textrm{UV}$ for different values of $L_1$ and $L_2$, with $\Delta E_\textrm{IR}$ fixed at 53~meV. (b) $\Delta t_\textrm{UV}$ and $\Delta E_\textrm{UV}$ for different values of $\Delta E_\textrm{IR}$ and $L_2$, with $L_1$ fixed at 100~$\mu$m. In both (a) and (b), $L_2$ ranges from 10 to 150~$\mu$m in steps of 5~$\mu$m. (c) Calculated $\Delta t_\textrm{UV}$ and $\Delta E_\textrm{UV}$ as a function of $L_1$ with fixed $\Delta E_\textrm{IR}$ and $L_2$. For small $L_1$, both bandwidth and pulse duration decrease with increasing $L_1$ due to the reduced bandwidth incident on the second BBO crystal. \tred{For sufficiently large $L_1$, $\Delta E_\textrm{UV}\Delta t_\textrm{UV}<1825$~meV~fs, indicating UV pulses with shapes characterized by lower limits on the FWHM time-bandwidth product than that for Gaussian pulses (Eq.~\ref{eq:unc})}. (d) Calculated $\Delta t_\textrm{UV}$ and $\Delta E_\textrm{UV}$ as a function of $\Delta E_\textrm{IR}$ with fixed $L_1$ and $L_2$. For small $\Delta E_\textrm{IR}$ the UV bandwidth is limited \tred{to twice the input bandwidth (dashed line)}; for large $\Delta E_\textrm{IR}$ the UV bandwidth is limited by the BBO phase matching bandwidth.} 
\label{fig:banana}
\end{figure*}

In Fig. \ref{fig:banana}(a) we plot $\Delta t_\textrm{UV}$ and $\Delta E_\textrm{UV}$ for various values of $L_1$ and $L_2$, with $\Delta E_\textrm{IR}$ fixed at 53~meV to match our experimental values. As discussed above, increasing $L$ decreases the phase matching bandwidth, which increases pulse duration. For larger values of $L_2$ the phase matching bandwidth is significantly exceeded, which is reflected by a deviation from the behavior of Fourier transform-limited Gaussian pulses (gray line). We also see that values of $L_2$ up to 25~$\mu$m are close to the Gaussian limit for all values of $L_1$ considered. This means that crystals thinner than 25~$\mu$m would provide only minimal resolution improvements over a 25~$\mu$m crystal. We also note that thinner free-standing BBO crystals are not commercially available \cite{Homann12}.

The behavior when independently varying $L_1$ is less straightforward because the resulting changes on the 3~eV pulse must then be propagated through the second BBO. \tred{Counter-intuitively, Fig.~\ref{fig:banana}(a) shows larger values of $L_1$ generate shorter UV pulses with decreased divergence from the Gaussian limit. To understand this behavior, in Fig.~\ref{fig:banana}(c)} we plot $\Delta t_\textrm{UV}$ and $\Delta E_\textrm{UV}$ as a function of $L_1$ for $L_2=50\mu$m. For $L_1 > 300\mu$m the behavior is as expected: decreasing bandwidth and increasing pulse duration with increasing thickness. Interestingly, below 300~$\mu$m the trend for the pulse duration is reversed. This behavior can be understood as follows: for sufficiently thin $L_1$, the bandwidth of the SHG generated in the first crystal will exceed the phase-matching bandwidth of the second crystal. This leads to non-Gaussian pulses in the second crystal which then deviate from the Fourier limit for Gaussian pulses. 

Now we fix $L_1$ at 100~$\mu$m, our experimental value, and vary $L_2$ and the bandwidth of the fundamental $\Delta E_\textrm{IR}$. In Fig. \ref{fig:banana}(b) we plot $\Delta t_\textrm{UV}$ and $\Delta E_\textrm{UV}$ for various values of $\Delta E_\textrm{IR}$ and $L_2$. Increasing $L_2$ improves energy resolution at the expense of time resolution, as is expected. As before, UV pulses diverge from the Fourier limit for large $L_2$.

To better illustrate the evolution from the thin-crystal to thick-crystal limit, in Fig. \ref{fig:banana}(d) we plot UV bandwidth and pulse duration as a function of $\Delta E_\textrm{IR}$. For small $\Delta E_\textrm{IR}$, where the \textit{thin crystal} limit applies, $\Delta E_\textrm{UV} \approx 2\Delta E_\textrm{IR}$. This is because $\mathcal{E}_\textrm{6eV}(t) \propto \mathcal{E}_\textrm{3eV}(t)^2 \propto \mathcal{E}_\textrm{1.5eV}(t)^4$, which leads to a doubling of the bandwidth of Gaussian pulses. As $\Delta E_\textrm{IR}$ increases, $\Delta E_\textrm{UV}$ asymptotically approaches the phase matching bandwidth of the BBO.

This allows us to identify two regimes for 6~eV generation. In the first, $\Delta E_\textrm{UV}$ is limited by $\Delta E_\textrm{IR}$; here decreasing $\Delta E_\textrm{IR}$ improves energy resolution and worsens time resolution. In the second regime, $\Delta E_\textrm{UV}$ is limited by the BBO phase matching bandwidth, and 6~eV pulses begin to become non-Gaussian. Here decreasing $\Delta E_\textrm{IR}$ improves both energy and time resolution. 

In practice, $\Delta E_\textrm{IR}$ is largely limited by fundamental properties of the driving laser system and difficult to change. When designing an experiment, time and energy resolutions are selected by choosing $L_1$ and $L_2$. The 6~eV generation has a smaller phase matching bandwidth, so $L_2$ is the more sensitive degree of freedom. It is reasonable to choose $L_2$ based on the target resolutions, then select $L_1$ such that $\Delta t_\textrm{UV}$ is optimized. We will further discuss strategies for optimizing a trARPES experiment in Section~\ref{s:Discussion}.

\subsection{Calculations in Experimental Conditions}

To compare to our measurements of time and energy resolution (see Section~\ref{s:Experiment}), we now evaluate Eq.~\ref{eq:shg} for our experimental parameters. We use $L_1=100~\mu$m, and $L_2\in\{25,50,100\}~\mu$m. We calculate the second and fourth harmonic spectra generated from an unchirped Gaussian fundamental pulse at 1.5~eV for these three specific sets of BBO thicknesses. The outputs of the calculation that we present are the spectral and temporal intensities ($\tilde I(\lambda)$ and $I(t) = \lvert \mathcal{E}(t) \rvert ^2$) of the second and fourth harmonic pulses.

\begin{figure}
\includegraphics[width=\columnwidth]{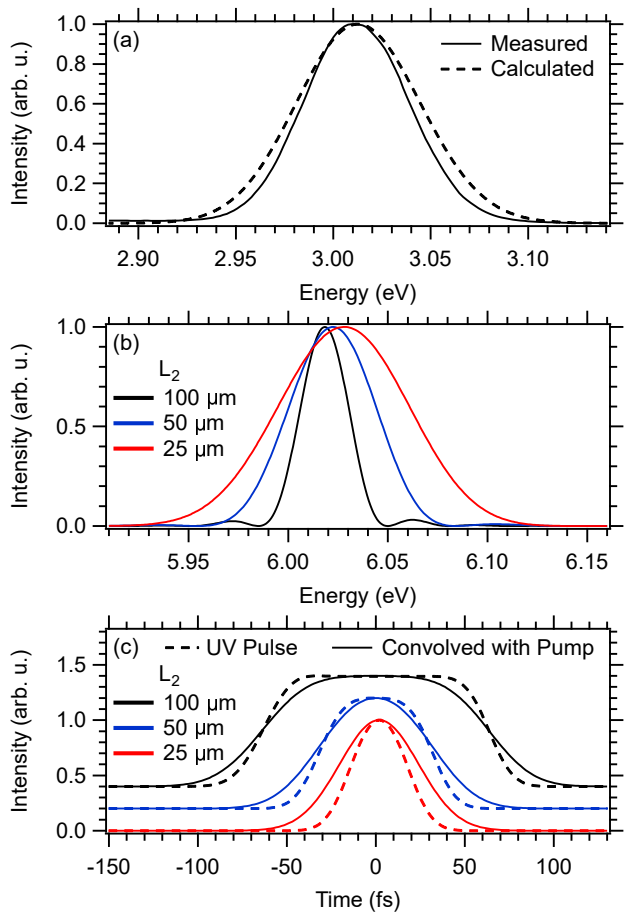}
\caption{(a) 3~eV spectra as measured (solid line) and calculated (dotted line) \tred{using $L_1$ = 100 $\mu$m}. Calculations approximately agree with data. (b) 6~eV spectral and (c) temporal intensity calculated using Eq. \ref{eq:shg} for different BBO thicknesses $L_2$ with $L_1$ fixed at 100~$\mu$m and $\Delta E_\textrm{IR}=53$~meV. In (c), the dotted and solid lines are before and after convolving with a 39~fs Gaussian pump pulse, respectively.}
\label{fig:thy}
\end{figure}

For each BBO thickness we first use the measured spectrum of the 1.5~eV fundamental to calculate the spectrum of the second harmonic near 3~eV. The comparison between measured and calculated spectra is shown in Fig.~\ref{fig:thy}(a), with overall agreement. Next, we propagate the calculated 3~eV spectra through another second harmonic calculation to arrive at the spectrum near 6~eV, shown in Fig.~\ref{fig:thy}(b). Note that we do not directly compare with any measured optical spectra here because our optical spectrometer lacks the necessary energy resolution at this wavelength; however, in Section~\ref{s:Experiment} we will compare with the experimentally more relevant  measurement of $\Delta E_{\textrm{UV}}$ from photoemission spectroscopy. In Fig.~\ref{fig:thy}(c) we further calculate the temporal intensity $I_{\textrm{6eV}}(t)$ as well as its convolution with $I_{\textrm{1.5eV}}(t)$ (Gaussian of width $\Delta t_\textrm{IR}=39$~fs), which we will compare with experimentally measured pump-probe cross-correlations.

\begin{table}
\begin{ruledtabular}
\begin{tabular}{l|lll}
& \multicolumn{3}{c}{Calculated Values}  \\
$n(\lambda)$ Source & $\Delta t_\textrm{UV}$ & $\Delta t_\textrm{Exp}$ & $\Delta E_\textrm{UV}$\\
\hline
$L_2=25\mu$m                &           &           &          \\
Castech                     & 26.1~fs   & 46.4~fs   & 76.7~meV \\
Tamosauskas \textit{et al.} & 26.9~fs   & 46.7~fs   & 75.2~meV \\
Eksma                       & 25.3~fs   & 46.1~fs   & 78.7~meV \\
\hline
$L_2=50\mu$m                &           &           &          \\
Castech                     & 52.3~fs   & 59.8~fs   & 50.6~meV \\
Tamosauskas \textit{et al.} & 54.9~fs   & 61.5~fs   & 49.0~meV \\
Eksma                       & 49.0~fs   & 57.8~fs   & 52.9~meV \\
\hline
$L_2=100\mu$m                &          &           &          \\
Castech                     & 115~fs    & 116~fs    & 27.8~meV \\
Tamosauskas \textit{et al.} & 120~fs    & 121~fs    & 26.7~meV \\
Eksma                       & 108~fs    & 109~fs    & 29.3~meV \\
\end{tabular}
\end{ruledtabular}
\caption{Calculated values for $\Delta t_\textrm{UV}$ and $\Delta E_\textrm{UV}$ using indices of refraction taken from three different sources.\cite{Castech,Tamosauskas2018,Eksma} Calculations are based on a fixed 53~meV bandwidth of the fundamental 1.5~eV pulse, $L_1=100~\mu$m and three values of $L_2$.}
\label{t:compareior} 
\end{table}

The fourth harmonic spectra show the expected spectral broadening for thinner crystals as the phase matching bandwidth increases. The shift to lower energies for thicker crystals reflects the rapid dropoff of SHG efficiency when approaching the type-I phase-matching limit of 6.05~eV photon energy in BBO \cite{Kato1986}. For $L_2=100~\mu$m we observe side peaks and a tendency toward a rectangular temporal pulse shape; as discussed above, these reflect the fact that the 3~eV bandwidth exceeds the phase matching bandwidth of this crystal thickness. This necessarily indicates a deviation from the Gaussian Fourier limit at this crystal thickness.

When obtaining calculated values for 6 eV bandwidth and pulse duration, we compare values for the index of refraction from three different sources \cite{Castech,Eksma,Tamosauskas2018}, see Table~\ref{t:compareior}. We used the variation between these three sources to estimate uncertainties for the calculated values of $\Delta E_\textrm{UV}$, $\Delta t_\textrm{UV}$, and, after convolving the 6 eV pulse shapes with the pump pulse, $\Delta t_\textrm{Exp}$. The values in Table~\ref{t:compareior} have been calculated for a fixed 53~meV bandwidth of the fundamental 1.5~eV pulse.

\section{Experiment}\label{s:Experiment}
\subsection{Experimental Setup}

Our experimental setup is depicted in Fig. \ref{fig:setup}. Our ultrafast laser source is based on a Coherent Vitara-T that seeds a RegA 9050 amplifier. It generates 1.5 eV pulses with a repetition rate of 312 kHz at $\sim$1.5 W ($\sim 5~\mu$J pulse energy). A commercial frequency-resolved optical grating (FROG) \cite{OShea2001} from Swamp Optics was used to characterize the 1.5~eV pulses to reveal a 36~fs pulse duration and 53~meV bandwidth, with a time-bandwidth product of 1908~meV~fs, or 105\% of the Gaussian Fourier limit. The fundamental beam is split into pump and probe paths. The pump beam is focused onto the sample after passing through a delay stage. The pump pulse duration was measured to be 39 fs at the sample position using the FROG.

The probe beam is frequency quadrupled to 6~eV in two BBO crystals of thickness $L_1$ and $L_2$. The pulses acquire chirp as they pass through dispersive media \cite{Weiner2009}. A prism compressor \cite{Fork1984} after each BBO introduces a negative GVD and thus minimizes the pulse duration at the 2nd BBO and the sample. By varying $L_2$ we tune the bandwidth and duration of the 6~eV pulses, subject to the lower bound set by the Gaussian Fourier limit in Eq. \ref{eq:unc}. As described above, we use $L_1=100~\mu$m and $L_2\in\{25,50,100\}~\mu$m. 25~$\mu$m represents the thinnest freestanding BBOs commercially available \cite{Homann12}. Thinner crystals must be supported by bonding to optical substrates, which can introduce additional issues with long-term stability, beam pointing and dispersion management.

After switching crystals we adjusted the angle of the BBO to maximize 6~eV intensity. We also optimized the GVD of the prisms, which must be adjusted whenever the amount of dispersive media in the beam path changes. The prism compressor after the first BBO controls the pulse duration of the intermediate 3~eV beam. Because of the nonlinear nature of the SHG process, the shortest 3~eV pulses correspond to the strongest 6~eV beam. This means the alignment of the first prism compressor can be optimized by maximizing the 6~eV photoemission intensity. The prism compressor after the second BBO was optimized by repeatedly measuring the time resolution.

\subsection{Resolution Measurements}

We use $\Delta t_\textrm{IR}$ and $\Delta t_\textrm{UV}$ to represent the FWHM of the temporal intensity of infrared pump and ultraviolet probe pulses, respectively. Assuming Gaussian pulses, the experimental time resolution is
\begin{equation}
\Delta t_\textrm{Exp} = \sqrt{\Delta t_\textrm{IR}^2+\Delta t_\textrm{UV}^2} \textrm{.}
\label{eq:res}
\end{equation} 
\noindent We also use $\Delta E_\textrm{IR}$ and $\Delta E_\textrm{UV}$ to represent the FWHM of the spectral intensity of pump and probe pulses, respectively. Excluding extrinsic factors such as space charge \cite{Hellmann2009} and photoelectron analyzer resolution, we consider $\Delta E_\textrm{UV}$ to be close to equal to our experimental energy resolution $\Delta E_\textrm{Exp}$. This is an assumption and not generally correct. In reality, the energy resolution can be dominated by space charge broadening when other experimental considerations such as limits on data acquisition time demand a higher photon flux. 

We characterized our experimental time resolution $\Delta t_\textrm{Exp}$ and energy resolution $\Delta E_\textrm{UV}$ by measuring Bi$_2$Se$_3$ and polycrystalline copper, respectively. Bi$_2$Se$_3$ samples were cleaved in vacuum. Copper was scratched in vacuum to expose a fresh surface. Photoelectrons were collected by a Scienta R4000 hemispherical analyzer at 85 K and pressures under 10$^{-10}$ Torr. Incident pump fluences were 300-500~$\mu$J/cm$^2$.

The energy resolution was obtained by measuring the Fermi edge of copper in equilibrium and fitting a Fermi-Dirac function with linear background convolved with a Gaussian of FWHM $\Delta E_\textrm{UV}$. The linear background accounts for the density of states. To ensure limited broadening from space charge effects, we took measurements as a function of decreasing UV flux and recorded the final dataset only once the fitted energy resolution stabilized at a minimum value. We estimate the errors of the energy resolution to be $\pm 2$~meV.

The time resolution was measured from the photoemission intensity $\sim1.5$~eV above the Fermi level of Bi$_2$Se$_3$ \cite{Sobota2012}. We fit the population dynamics of a subregion of the unoccupied states to a Gaussian with FWHM $\Delta t_\textrm{Exp}$ convolved with two exponential response functions $\sum_{i=1,2} e^{-(t-t_0)/\tau_i}\Theta(t-t_0)$ where $\Theta(t)$ is the Heaviside step function. \tred{The double exponential decay fit was chosen to match the data, and accounts for multiple electron decay channels.} Although the assumption of Gaussian pulses may not be justified for larger $L_2$, for consistency we fit to the same function for all three values of $L_2$.

\begin{figure}
\includegraphics[width=\columnwidth]{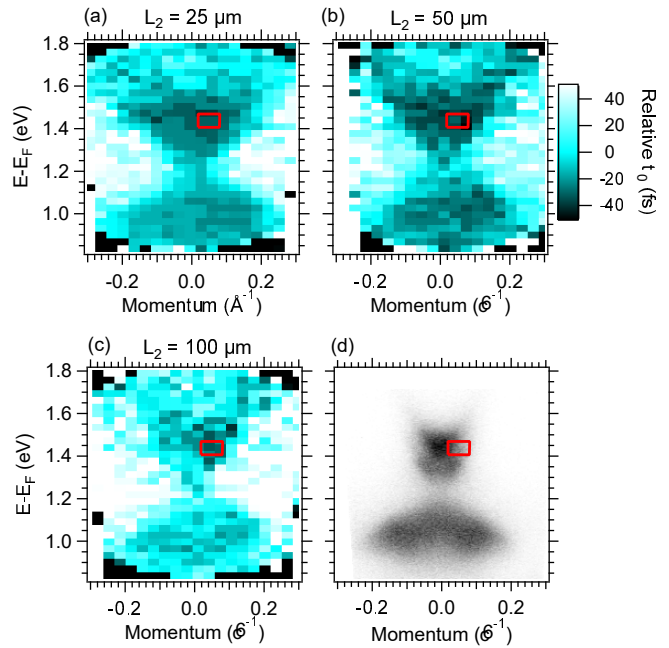}
\caption{ (a-c). $t_0(E,k)$ for the three values of $L_2$ used. \tred{Zero is arbitrarily chosen as the average value of $t_0$ for each panel.} The red square is the integration region used for time resolution fits. It is located in the area with earliest $t_0$, and thus fastest rise time. (d). Transient photoemission intensity over the same window as (a-c). }
\label{fig:timemapping}
\end{figure}

We select the fitting region, depicted by a red square in Fig. \ref{fig:timemapping}, using the time mapping procedure described elsewhere \cite{Mathias2010,Soifer2019} to identify the area with the fastest rise time. For the purpose of the time mapping procedure, we define $t_0(E,k)$ to be the delay at which photoemission intensity is maximal for a specific $(E,k)$ point in the excited band structure. Smaller values of $t_0$ represent faster rise times. We plot $t_0(E,k)$ for $L_2\in\{25,50,100\}~\mu$m in Fig.~\ref{fig:timemapping}(a-c). The fitting region has the earliest $t_0$, and thus the fastest rise time, indicating that is populated by direct optical transitions rather than secondary scattering processes which obscure the actual time resolution  \cite{Weinelt2004,Reimann2014}.

A large energy range data set of the dynamics on Bi$_2$Se$_3$ is presented in a Supplemental Movie. The first image potential state is visible at negative delays; these electrons are photoemitted through multi-photon processes \cite{Sobota2012}.

\subsection{Results}

\begin{figure*}
\includegraphics[width=\textwidth]{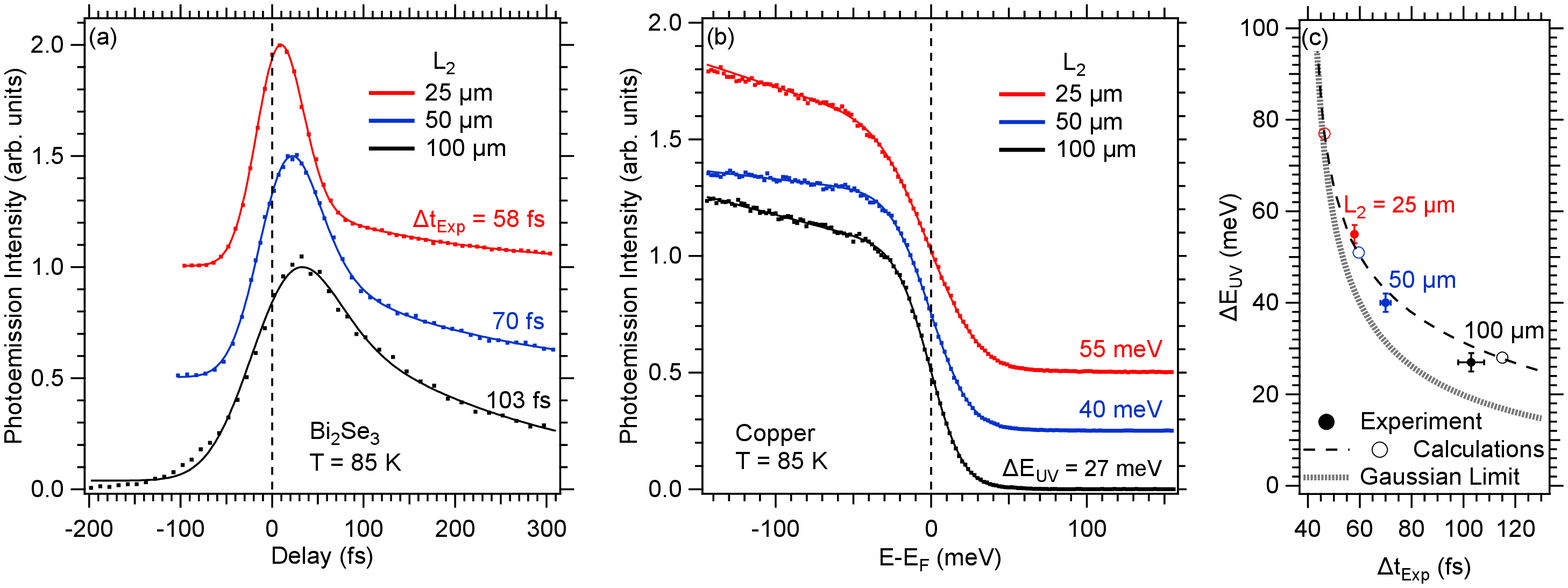}
\caption{(a) Photoemission intensity vs. delay for three BBO thicknesses integrated over the region indicated by the red square in Fig.~\ref{fig:timemapping}. Dots are data, solid lines are fits to a Gaussian convolved with two exponential decay functions. (b) Fermi edge of copper for three BBO thicknesses.  Dots are data, solid lines are fits to a Fermi-Dirac convolved with energy resolution and with linear background added to account for the density of states in Cu. (c) Experimental (data points with error bars) and calculated (dashed line and circles) time and energy resolutions for several BBO thicknesses. The thick dotted line is the Gaussian Fourier limit based on $\Delta t_\textrm{IR}=39$~fs pump pulses in our experiment.}
\label{fig:data}
\end{figure*}

Experimental data and corresponding fits are shown in Fig.~\ref{fig:data}. Our measured time and energy resolutions are reported in Table~\ref{t:summary}.  The best achieved time resolution is 58 $\pm$ 1~fs. The general trend is that thinner crystals have better time resolution, and thicker crystals have better energy resolution, as expected. 

\begin{table*}
\begin{ruledtabular}
\begin{tabular}{l|llll|llll}
            & \multicolumn{4}{c}{Experimental Values}                                                                                      & \multicolumn{4}{c}{Calculated Values}\\
BBO $L_2$ & $\Delta t_\textrm{Exp}$ & $\Delta t_\textrm{UV}$ & $\Delta E_\textrm{UV}$ & $\Delta E_\textrm{UV}\Delta t_\textrm{UV}$ & $\Delta t_\textrm{Exp}$ & $\Delta t_\textrm{UV}$   & $\Delta E_\textrm{UV}$ & $\Delta E_\textrm{UV}\Delta t_\textrm{UV}$\\
\hline
25~$\mu$m & 58 $\pm$ 1~fs  & 43 $\pm$ 3~fs  & 55 $\pm$ 2~meV  & 2360~meV~fs (129\%) & 46 $\pm$ 1 fs  & 26 $\pm$ 1 fs   & 77 $\pm$ 2 meV    & 2000~meV~fs (110\%) \\
50~$\mu$m & 70 $\pm$ 2~fs  & 58 $\pm$ 5~fs  & 40 $\pm$ 2~meV & 2320~meV~fs (127\%)  & 60 $\pm$ 2 fs  & 52 $\pm$ 3 fs   & 51 $\pm$ 2 meV    & 2650~meV~fs (145\%) \\
100~$\mu$m& 103 $\pm$ 5~fs & 95 $\pm$ 10~fs & 27 $\pm$ 2~meV & 2570~meV~fs (141\%)  & 115 $\pm$ 6 fs & 114 $\pm$ 6 fs & 28 $\pm$ 1 meV    & 3190~meV~fs (175\%)

\end{tabular}
\end{ruledtabular}
\caption{Comparing measured time and energy resolutions to calculations. In all cases $L_1=100$~$\mu$m. The experimental durations of the 6~eV probe pulse $\Delta t_\textrm{UV}$ are calculated using Eq.~\ref{eq:res} for a fixed pump pulse duration $\Delta t_\textrm{IR}=39$~fs. Calculations are based on a fixed 53~meV bandwidth of the fundamental 1.5~eV pulse. Error bars for calculations are obtained by comparing results for different sources for $n(\lambda)$ \cite{Kato1986,Castech,Eksma}, see Table~\ref{t:compareior}.}
\label{t:summary}
\end{table*}

We also compare the experimental and calculated results in Table~\ref{t:summary} and Fig.~\ref{fig:data}(c). Calculated values are extracted from the curves in Fig.~\ref{fig:thy}(b-c). Overall, the calculated resolutions qualitatively agree with the experimental trend, and quantitatively agree within 30\%. Discrepancies may be largely attributable to uncertainty in the precise value of $L_2$. For instance, calculated values for $L_2$ = 35 $\mu$m are $\Delta t_\textrm{UV}$ = 36 fs and $\Delta E_\textrm{UV}$ = 63 meV, a much closer agreement to the experimental values for our thinnest crystal.

Table~\ref{t:summary} also reports the product of our probe pulse length and bandwidth as the time-bandwidth product and, in parentheses, how closely it approaches the ideal value for Gaussian pulses. The experimental probe pulse length $\Delta t_\textrm{UV}$ is calculated from the measured value $\Delta t_\textrm{Exp}$ using Eq.~\ref{eq:res}, with $\Delta t_\textrm{IR}=39$~fs. With increasing BBO thickness $L_2$ both experimental and calculated values depart from the Gaussian Fourier limit and approach the larger limit for a rectangular pulse.

\section{Discussion}\label{s:Discussion}

We proceed to discuss practical ways to control the partitioning of the time-bandwidth product in trARPES experiments as suggested by our results. Ideal probe pulses should be Gaussian, in which case the Gaussian Fourier limit can be maintained. Target time and energy resolutions can be estimated using Eq.~\ref{eq:unc} and Eq.~\ref{eq:res}. Gaussian probe pulses require nonlinear crystals which are thin enough for the phase matching bandwidth to significantly exceed the desired energy resolution. In this \textit{thin crystal} limit, each SHG process increases bandwidth by a factor of $\sqrt{2}$. For a setup with two stages of SHG, this means the bandwidth of the fundamental should be half the desired energy resolution.

In practice, fine control of the fundamental bandwidth is often not possible. In this case, target time and energy resolutions may still be achieved by selecting proper nonlinear crystals. If the desired energy resolution is close to half the bandwidth of the fundamental, the approach described above may be followed. This will also give the best possible time resolution for the laser source. If better energy resolution, and hence worse time resolution, is desired, thicker crystals must be used. This necessarily implies deviation of the time-bandwidth product from the Gaussian Fourier limit.

For a setup with two stages of SHG, our calculations in Fig.~\ref{fig:shg} and \ref{fig:banana} show that 6~eV bandwidth and pulse duration are affected far more by the thickness of the second nonlinear crystal $L_2$ than the first crystal $L_1$. Therefore it is advisable to first choose this crystal based on the desired resolutions and input bandwidth. Fig.~\ref{fig:banana}(b) may be used as a guide. The thickness of the first crystal $L_1$ may then be selected to minimize the 6~eV pulse duration, as demonstrated in Fig.~\ref{fig:banana}(c), resulting in a modest improvement in both time and energy resolution over the na\"ive approach of choosing the thinnest possible crystal. 

Beyond approaching the Fourier limit most closely for a given input bandwidth, this strategy may also offer the additional advantage of maximizing the 6~eV flux for a given set of resolutions. To give an example for our setup, Fig.~\ref{fig:banana}(c) shows that $L_1=100~\mu$m is too small for optimally approaching the Gaussian Fourier limit using $L_2=50~\mu$m. For $L_1=100~\mu$m, the calculated 6~eV pulse duration is 52~fs, while going to $L_1=300~\mu$m would have enabled a marginal improvement to 50~fs \tred{as well as a significant increase in photoemission intensity. As Eq.~\ref{eq:shg} shows, the intensity of each SHG process} scales with $L^2$ such that a $3\times$ thicker BBO $L_1$ gives a $9\times$ increased 3~eV flux. In turn this increases the 6~eV flux, \tred{and thus photoemission intensity}, considerably by $81\times$ if all other parameters remain unchanged. This can be an important practical consideration when trying to obtain the best time resolution as this requires smaller $L_2$ values that decrease the 6~eV flux quadratically, \tred{as well as for photoemission setups where photon flux is a limiting factor}.

\begin{figure}[t]
\includegraphics[width=\columnwidth]{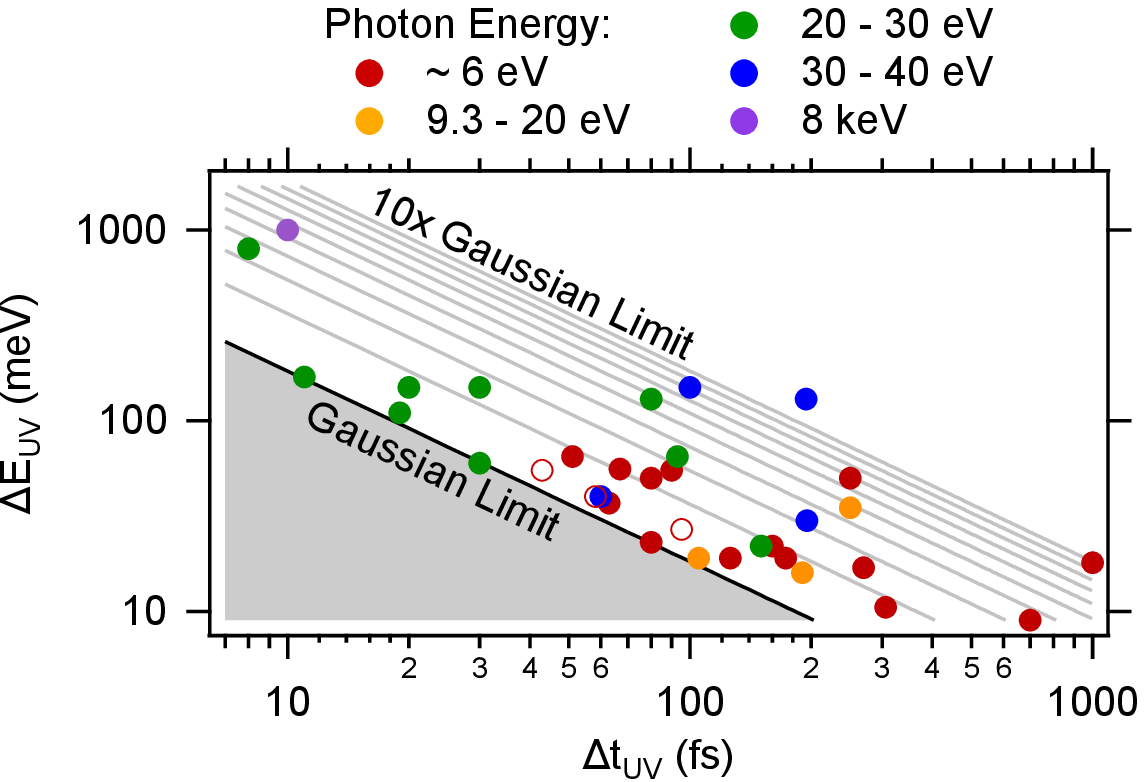}
\caption{Comparing probe pulse duration and bandwidth from various reports of time-resolved photoemission. Black line is Gaussian Fourier limit 1825~meV~fs. Gray lines are integer multiples of Gaussian limit $1825n$~meV~fs up to $n=10$. Values reported here are represented by open circles. Solid points are from papers in Table \ref{t:groups}. }
\label{fig:compare}
\end{figure}

For completeness, we acknowledge limitations in our calculations which neglected (1) \textit{pump depletion}, (2) \textit{group velocity dispersion}, and (3) \textit{geometric focusing effects}, all of which can be relevant factors in a real SHG process. These can be accounted for by numerically solving a pair of differential equations which account for the nonlinear interactions of fundamental and second harmonic as a function of time and space  \cite{Sidick_1995_ultrashort,Sidick_1995_ultrashort_2}. While a complete treatment is outside the scope of this work, here we highlight some of the salient conclusions for practitioners:

(1) \textit{Pump depletion} refers to the situation in which the fundamental pulse loses substantial energy due to generation of the second harmonic, and is thus associated with saturation of the second-order response. It manifests as a temporal asymmetry in the shape of the second harmonic pulse \cite{Sidick_1995_ultrashort}, and can therefore be an important factor to consider if asymmetric cross-correlations are observed in a trARPES measurement. 

(2) \textit{Group velocity dispersion} in the fundamental will temporally broaden the pulse and reduce conversion efficiency accordingly, though \textit{a priori} it is unclear whether this broadening can be compensated after the SHG process. When considering second-order chirp only, numerical calculations show that the ultimate shape and duration of the second harmonic pulse is independent of whether compression is performed on the fundamental or the second harmonic pulse \cite{Sidick_1995_ultrashort_2}. 

(3) \textit{Geometric focusing effects} are important because the majority of SHG will occur within the volume of the crystal defined by the beam waist and Rayleigh length. This consideration leads to well-defined criteria for focusing conditions which optimize overall conversion efficiency for a given crystal thickness \cite{Guha_1980_the}; pragmatically, this implies that one can design the focusing conditions after suitable crystal thicknesses are chosen according to the time-bandwidth criteria discussed in this work. Finally, we note that fundamental and second harmonic beams do not strictly co-propagate, but actually deviate by a finite \textit{walk-off angle}. In substantially thick crystals, this leads to spatial elongation of the second harmonic in the walk-off direction \cite{Weiner2009} which can compromise the ability to achieve a well-defined focus spot at the sample position.

\begin{table*}
\begin{ruledtabular}
\begin{tabular}{llllllll}
Reference                                           & Photon E.  & Rep. Rate    & $\Delta E_\textrm{Exp}$ & $\Delta t_\textrm{Exp}$ & $\Delta E_\textrm{UV}$ & $\Delta t_\textrm{UV}$ & \%\\
\hline
\textbf{Crystals} \\
Kuroda \textit{et al.}\cite{Kuroda2017}             & 5.16~eV    & 100~kHz      & 23~meV\cite{Guedde2019-pc}  & 113~fs  & 23~meV\cite{Guedde2019-pc} & 80~fs & 101\% \\
Ishida, \textit{et al.}\cite{Ishida2016}            & 5.9~eV     & 250~kHz      & 10.5~meV  & 350~fs  & 10.5~meV & 306~fs  & 176\% \\
Smallwood \textit{et al.}\cite{Smallwood2012-rsi}   & 5.93~eV    & $54.3/n$~MHz & 23~meV    & 310~fs  & 17~meV   & 270~fs  & 252\% \\
Freutel \textit{et al.}\cite{Freutel2019}           & 6.0~eV     & 250~kHz      & 55~meV    & 100~fs  & 55~meV   & 90~fs   & 271\% \\
Monney \textit{et al.}\cite{Monney2018}             & 6.0~eV     & 300~kHz      & 50~meV    & 250~fs  & 50~meV   & 250~fs  & 685\% \\
Sobota \textit{et al.}\cite{Sobota2012}             & 6.0~eV     & 80~MHz       & 22~meV    & 163~fs  & 22~meV   & 160~fs  & 193\% \\
This work                                           & 6.0~eV     & 312~kHz      & 55~meV    & 58~fs   & 55~meV   & 43~fs   & 129\% \\
                                                    & 6.0~eV     & 312~kHz      & 40~meV    & 70~fs   & 40~meV   & 58~fs   & 127\% \\
                                                    & 6.0~eV     & 312~kHz      & 27~meV    & 103~fs  & 27~meV   & 95~fs   & 141\% \\
Yang \textit{et al.}\cite{Yang2019-rsi}             & 6.05~eV    & 500~kHz      & 19~meV    & 130~fs  & 19~meV   & 126~fs  & 131\% \\
Boschini \textit{et al.} (2014)\cite{Boschini2014}  & 6.05~eV    & 100~kHz      & 50~meV    & 85~fs   & 50~meV   & 80~fs   & 219\% \\
Boschini \textit{et al.} (2018)\cite{Boschini2018}  &6.2~eV      & 250~kHz      & 19~meV    & 250~fs  & 19~meV   & 173~fs  & 180\% \\
Wegkamp \textit{et al.}\cite{Wegkamp2014}           & 6.19~eV    & 40~kHz       & 90~meV    & 89~fs   & 37~meV\cite{Mor2017}& 63~fs   & 128\% \\
Faure \textit{et al.}\cite{Faure2012}               & 6.28~eV    & 250~kHz      & 70~meV    & 76~fs   & 56~meV   & 67~fs   & 206\% \\
Parham \textit{et al.}\cite{Parham2017}             & 6.28~eV    & 20~kHz       & 9~meV     & 700~fs  & 9~meV    & 700~fs  & 345\% \\
Andres \textit{et al.}\cite{Andres2015}             & 6.3~eV     & 300~kHz      & 65~meV    & 70~fs   & 65~meV   & 51~fs   & 182\% \\
Yang \textit{et al.}\cite{Yang2019-rsi}             & 6.7~eV     & 500~kHz      & 18~meV    & 1000~fs & 18~meV   & 1000~fs & 986\% \\
\hline
\textbf{Gases} \\
Cilento \textit{et al.}\cite{Cilento2016}           & 9.3~eV     & 250~kHz      & 100~meV   & 260~fs  & 35~meV   & 250~fs  & 479\% \\
Haight \textit{et al.}\cite{Haight1988}             & 10.7~eV *  & 200 Hz       & 100~meV   & $\sim$ps&95~meV&$\sim$ps     &       \\
Lee \textit{et al.}\cite{Lee2019}                   & 11~eV      & 250~kHz      & 16~meV    & 250~fs  & 16~meV   & 190~fs  & 162\% \\
Cucini \textit{et al.}\cite{Cucini2019}             & 16.9~eV *  & 250~kHz      & 22~meV    & 300~fs  & 19~meV   & 105~fs  & 109\% \\
Petersen \textit{et al.}\cite{Peterson2011}         & 20.4~eV *  & 1~kHz        & 150~meV   & 30~fs   & 150~meV  & 30~fs   & 247\% \\
Puppin \textit{et al.}\cite{Puppin2019}             & 21.7~eV    & 500~kHz      & 121~meV   & 37~fs   & 110~meV  & 19~fs   & 115\% \\
Rohde \textit{et al.}\cite{Rohde2016}               & 22.1~eV    & 10~kHz       & 185~meV   & 13~fs   & 170~meV  & 11~fs   & 102\% \\
Buss \textit{et al.}\cite{Buss2019}                 & 22.3~eV    & 50~kHz       & 60~meV    & 65~fs   & 60~meV   & 30~fs   & 99\%  \\
Eich \textit{et al.}\cite{Eich2014}                 & 22.3~eV    & 10~kHz       & 150~meV   & 38~fs   & 150~meV  & 20~fs   & 164\% \\
Wallauer \textit{et al.}\cite{Wallauer2016}         & 23.25~eV   & 100~kHz      &150~meV\cite{Guedde2019-pc}&80~fs&130~meV\cite{Guedde2019-pc}&80~fs&570\%\\
Mills \textit{et al.}\cite{Mills2019}               & 25~eV *    & 60~MHz       & 22~meV    & 190~fs  & 22~meV   & 150~fs  & 181\% \\
Corder \textit{et al.}\cite{Corder2018}             & 29.9~eV\cite{Allison2019-pc} *&88~MHz&110~meV& 181~fs&65~meV&93~fs&331\% \\
Gierz \textit{et al.}\cite{Gierz2015}        & 30 eV *    & 1 kHz        & 800 meV   & 13 fs   & 800 meV  & 8 fs    & 351\% \\
Sie \textit{et al.}\cite{Sie2019}                   & 33~eV *    & 30~kHz       & 30~meV    & 200~fs  & 30~meV   & 195~fs  & 321\% \\
Frietsch \textit{et al.}\cite{Frietsch2013}         & 32 \& 36~eV *&10~kHz      & 260~meV   & 125~fs  & 150~meV  & 100~fs  & 822\% \\
Roth \textit{et al.}\cite{Roth2019}                 & 37~eV *    & 6~kHz        & 130~meV   & 200~fs  & 130~meV  & 194~fs  & 1380\%\\
\hline
\textbf{X-Ray / FEL} \\
Kutnyakhov \textit{et al.}\cite{Kutnyakhov2019}      & 36.5~eV *  & 5 kHZ& 130~meV\cite{Kutnyakhov2019-pc}& 150~fs& 40~meV   & 60~fs   & 132\% \\
Oloff \textit{et al.}\cite{Oloff2016}           & 8 keV *     & 30 Hz        & 1.25~eV   & 40~fs   & 1~eV     & 10~fs   & 548\% \\
\end{tabular}
\end{ruledtabular}
\caption{Comparing time and energy resolutions of various reports of time-resolved photoemission. Photon energies listed are those used for the measurements; a star ($*$) indicates significant tunability of photon energy. $\Delta E_\textrm{Exp}$ and $\Delta t_\textrm{Exp}$ are achieved experimental energy and time resolutions, including effects such as space charge and analyzer resolution, but perhaps in a low flux limit. In most cases, experimental resolutions are practically achievable in day-to-day operations. $\Delta E_\textrm{UV}$ and $\Delta t_\textrm{UV}$ are the reported duration and bandwidth of the probe pulse only, some of which are conservative estimates. These were calculated by subtracting $\Delta t_\textrm{IR}$ and analyzer energy resolution from experimental resolutions in quadrature, when available. If no data on probe pulse duration or analyzer resolution is given we assume the UV pulse duration and bandwidth are same as experimental resolution. $\% = \Delta t_\textrm{UV}\Delta E_\textrm{UV}/1825$~meV~fs is how closely the probe pulse approaches the Gaussian Fourier limit. Corresponding authors were consulted to ensure accurate values. }
\label{t:groups}
\end{table*}

Finally, we compile the recent trARPES literature with an emphasis on time and energy resolution in Table~\ref{t:groups} and Fig.~\ref{fig:compare}. Probe photon energy and laser repetition rates are equally important experimental considerations and are included as well. We consider trARPES instruments based on frequency upconversion in nonlinear crystals, HHG in gases and recently emerging free electron laser sources. This compilation aims at highlighting the diversity and specialization the field has achieved in the last decade. 

\tred{We emphasize that the time-bandwidth product should not be taken as the sole metric for evaluating a setup, especially since non-Gaussian pulse shapes can achieve lower limits \cite{Rulliere2005}. Nevertheless,} this summary allows us to identify general trends. First, we note that both HHG-based and crystal-based implementations routinely reach the Gaussian Fourier limit within a factor of 1.2 to 4. The most striking difference is the superior time resolution that most HHG-based instruments offer; probe pulse durations are commonly below 50~fs, with some below 10~fs. In contrast, crystal-based trARPES instruments operate with $>50$~fs time resolution, and often considerably slower, for the sake of energy resolution. Some crystal based instruments are able to achieve $<10$~meV energy resolution. Notably, achieving the best time resolution also requires very short pump pulses \cite{Gierz2017}, which can become a limiting factor when aiming for $<30$~fs time resolution. As far as the first pioneering publications suggest \cite{Oloff2016,Kutnyakhov2019}, trARPES at free electron lasers has the potential for excellent time resolution but mitigating space charge related degradation of spectral resolution will be a challenge. 

In summary, we tune the time and energy resolutions of a laser-based time-resolved ARPES setup by varying the thickness of the nonlinear crystals generating ultraviolet pulses. We report time and energy resolutions over the ranges of 58 - 103~fs and 55 - 27~meV. Calculations suggest a method to select crystal thickness for two-stage 6 eV generation based on desired experimental resolutions. The second nonlinear crystal should be selected such that its phase matching bandwidth matches the desired resolutions. Then the first crystal can be selected to optimize the experimental time resolution and probe intensity. This ability to measure at various time and energy resolutions is key to enabling diverse time-resolved photoemission measurements.

\begin{acknowledgments}
We gratefully acknowledge the corresponding authors who helped compile Table~\ref{t:groups}. This work was supported by the U.S. Department of Energy, Office of Science, Basic Energy Sciences, Materials Sciences and Engineering Division under contract No. DE-AC02-76SF00515. A.~G. acknowledges support from the Stanford Graduate Fellowship. N.~G. acknowledges support from the Swiss National Science Foundation under reference P2EZP2 178542. H.~P. acknowledges support from the German Science Foundation (DFG) under reference PF 947/1-1 and from the Advanced Light Source funded through the U.S. Department of Energy, Office of Science under contract No. DE-AC02-05CH11231.
\end{acknowledgments}

\section*{Data Availability} 

The data that support the findings of this study \tred{are} openly available in the Stanford Digital Repository \tred{at http://purl.stanford.edu/bz714qy3272 (http://doi.org/10.25740/bz714qy3272)}.

\bibliographystyle{apsrev4-1}
\bibliography{references}

\end{document}